\documentclass[amssymb,floats,tighten,prl,aps,epsf,enumerate]{revtex4}

\input{epsf.sty}

\draft

\def\be{\begin{equation}}
\def\ee{\end{equation}}

\begin{document}


\title{Magnetic collapse of a neutron gas: Can magnetars indeed be formed?}

\author{A. P\'erez Mart\'{\i}nez$^{a}$, H. P\'erez Rojas$^{a,b,c}$
and H. J. Mosquera Cuesta$^{c,d}$}

\address{\vskip 1cm $^a
$ Grupo de F\'{\i}sica Te\'orica,
ICIMAF, Calle E No. 309, 10400 La Habana, Cuba\\ $^b$ High Energy
Physics Division, Department of Physics, University of Helsinki
and\\ Helsinki Institute of Physics, P. O. Box 64, FIN-00014
Helsinki, Finland\\ $^c$ Abdus Salam International Centre for
Theoretical Physics, P.O. Box 586, Strada Costiera 11, Miramare
34100, Trieste, Italy \\ $d$ Centro Brasileiro de Pesquisas
F\'{\i}sicas, Laborat\'orio de Cosmologia e F\'{\i}sica
Experimental de Altas Energias \\ Rua Dr. Xavier Sigaud 150, CEP
22290-180, Urca, Rio de Janeiro, RJ, Brazil}

\begin{abstract}
A relativistic degenerate neutron gas in equilibrium with a
background of electrons and protons in a magnetic field exerts its
pressure anisotropically, having a smaller value perpendicular
than along the magnetic field. For critical fields the magnetic
pressure may produce the vanishing of the equatorial pressure of
the neutron gas. Taking it as a model for neutron stars, the
outcome could be a transverse collapse of the star. This fixes a
limit to the fields to be observable in stable neutron star pulsars as a
function of their density. The final structure left over after the
implosion might be a mixed phase of nucleons and meson condensate,
a strange star, or a highly distorted black hole or black "cigar",
but no any magnetar, if viewed as a super strongly magnetized
neutron star. However, we do not exclude the possibility of a superstrong
magnetic fields arising in supernova explosions which lead directly to
strange
stars. In other words, if any magnetars exist, they cannot be neutron
stars.
 \end{abstract}



\maketitle


\section{Introduction}

Gravitational collapse occurs in a body of mass $M$ and radius $R$
when its rest energy is of the same order of its gravitational
energy, i. e., $Mc^2 \sim G M^2/R$. We would like to argue that
for a macroscopic magnetized body, e.g., composed by neutrons in
an external field $\vec{B}$ ($|\vec{B}| \equiv B$), new physics
arises and a sort of collapse occurs when its internal energy
density $U$ is of the same order than its magnetic energy density
$\cal{\vec{M} } \cdot \vec{B}$, where $\vec{\cal M}$ is the
magnetization. This problem is interesting in the context of both
cosmology and astrophysics, as for instance in the study of
objects such as neutron stars. A gas of neutral particles having
an anomalous magnetic moment (as a model for neutron stars (NSs);
here we assume, as usually, a background of electrons and protons
in $\beta$ equilibrium,~ which is demanded by Pauli's Principle to
guarantee
neutron stability),
when placed in extremely strong magnetic fields has a nonlinear
(ferromagnetic) response to the external field and is also
unstable due to the vanishing of the transverse pressure for
surface fields strong enough ($B_{surf} \geq 10^{16}$ G). In this
phenomenon quantum effects play an essential role due to the
coupling of the particles' spins to the microscopic field
$\vec{B}$ seen by the particles (spin-polarization). For fields of
this order of magnitude there are values of the density for which
the magnetic energy of the system becomes of the same order of
magnitude than its total energy. At these physical conditions any
structure of super dense matter composed of neutral particles
having a magnetic moment may undergo a transverse collapse when
its pressure perpendicular to $\vec{B}$ vanishes. This implosion
is driven by the same mechanism described in \cite{Chaichian} for
charged particles.

We present in this new paper, which is a more elaborated version
of \cite{Hugo}, the main ideas concerning the role of ultra strong
magnetic fields in a gas of neutral particles: The standard model
of NSs, as envisioned by Duncan and Thompson in their model of
magnetars\cite{DT92}. The fundamental result that we obtain shows
that a NS, i. e., a neutron gas permeated by a super strong
magnetic field is unstable and must collapse. This inedit result
seems to ban the possibility of formation magnetars. We stress in
this respect the fact that similar results were obtained by two
different groups: Khalilov\cite{Khalilov} and Ghosh, Mandal and
Chakrabarty\cite{Chakrabarty1}. Although both teams of researchers
arrived to the same conclusion like ours, quite different
approaches were pursued.

The paper is organized as follows: Section II reviews the concept
of anisotropic pressure in self-gravitating systems like NSs. In
Section III are given the tools to construct the energy-momentum
tensor of a neutron gas. Section IV discusses the main differences
between the classical and quantum collapse of a gas configuration
in an approximation-independent way, based on the sign of the
electromagnetic response of the medium to the external applied
field. In Section V the dynamics of the neutron gas composing a NS
is studied in the presence of a magnetic field, followed in Section
VI by the derivation of its thermodynamical potential and
magnetization. Section VII discusses the conditions for the
collapse to take place, and an application of this physics to the
stability analysis of the proposed magnetars is given. The main
result of this paper shows that such objects should not form if
they are envisioned as a standard neutron gas in Fermi beta
equilibrium as is claimed in the original idea introducing the
 concept of magnetar. Some closing remarks and further potential
applications of this theory are part of the final section.

\section{Anisotropic pressures in self-gravitating systems}

The issue of local anisotropy in pressures was extensively
reviewed by Herrera and Santos\cite{Herrera} in a general
relativistic approach. These authors present several physical
mechanisms for its origin in both extremely low and very high
density systems, which may include astrophysical compact objects.
In the case of highly dense systems, it  was pointed out that
``exotic" phase transitions could occur during gravitational
collapse, which is exactly the problem we are concerned with in
this paper. However, the more fundamental problem regarding the
appearance of anisotropic pressures in strongly magnetized compact
stars was left open. More recently, Mak and Harko \cite{Mak}
present a class of exact solutions of Einstein's equations
corresponding to anisotropic stellar configurations which  can
describe realistic neutron stars.

We want to provide a more detailed description of the arising of
anisotropic pressures in a relativistic system like a NS,  an
essential point in understanding the physics behind the problem of
stability of ultra magnetized compact stars. We shall give firstly
general arguments to support this our view, and then we
concentrate in the specific calculations, in the one-loop
approximation, of the thermodynamic potential of the neutron star
configuration and its magnetization, the properties upon which the
most crucial conclusions can be drawn.

To fix ideas, we shall work in the grand canonical ensemble, and
we are considering as subsystem, some region inside the star. Such
subsystem is under the influence of the magnetic field created by
the rest of the system, which we name $\vec{H}$ ($|\vec{H}| \equiv
H $). The response of the subsystem is to polarize itself creating
a magnetization ${\cal \vec{M}}$ in the medium (the neutron gas)
satisfying the relation: $\vec{H}=\vec{B}-4\pi{\cal{\vec M}}$.
Obviously, inside the subsystem the microscopic field is
$\vec{B}=\vec{H}+4\pi{\cal {\vec M}}$, since $\vec{B}$ (named also
magnetic induction) and the electric field $\vec{E}$ are the
\textit{true fields} acting on the electric charges and magnetic
dipoles of elementary particles \cite{Sommerfeld} \cite{Jackson}.
(Note, however, that Landau in Ref.\cite{Landau3} uses the notation
$\vec{H}$ to name the magnetic field in vacuum). The field
$\vec{B}$, as well as $\vec{E}$, satisfy the Maxwell equations for
particles in vacuum. In what follows, when we are to discuss the
dynamics of the particles in the neutron gas we will sometimes
refer to $\vec{B}$ as the \textit{external magnetic field}, as is
usually named in quantum field theory and astrophysics. For an
external distant observer, $\vec{B}=\vec{H}$ (in Gaussian units)
since the magnetization is assumed to exist only inside the star.
We emphasize that actually $\vec{B}$ and $\vec{H}$ are external fields
within different contexts: $\vec{H}$ is external to the {\it subsystem}
object of study in the grand canonical ensemble, whereas $\vec{B}$ is
external to any particle chosen in the subsystem (it feels, in addition
to $\vec{H}$, the contribution from the magnetization field $4\pi{\cal
{\vec M}}$ due to the other particles of the subsystem).

In the case of a gas of electrically charged particles in an external
constant magnetic field $\vec{B}$, in classical electrodynamics, it is
the Lorentz force ${\vec F}=e{\vec v} \times {\vec B}/c$ the source of
an asymmetry in the pressure components parallel and perpendicular to
${\vec B}$. By writing $e{\vec v}={\vec j}\Delta V$, where $\Delta V=
dx_1 dx_2 dx_3$, calling $f_i = {F}_i/\Delta V$ as the $i$-th component
of the force density, and substituting ${\vec j} = c {\nabla}\times
{\cal {\vec M}}$, one has

\begin{equation}
f_i = -(\partial_i {\cal M}_s) B_s + (\partial_s {\cal M}_i) B_s
\label{LFD} \;.
\end{equation}

By multiplying by $\Delta V= dx_1 dx_2 dx_3$ and assuming $B_s=
B\delta_{s3}$ and $\partial{\cal M}_i/\partial x_3=0$ (actually it is
also ${\cal M}_i= {\cal M}\delta_{i3}$), only the first term in
(\ref{LFD}) remains nonzero and one recovers an expression for the
Lorentz force, which is obviously perpendicular to the field
$\vec{B}$.  For the corresponding pressure it yields $P_{L\perp}= -
{\cal {\vec M}} \cdot \vec{B}$. This is a classical effect and
obviously $P_{L\perp}$ must be added to the usual kinetic isotropic
pressure $P_0$, so that the total transverse pressure becomes
$P_{\perp}=P_0 + P_{L\perp}$. As in classical electrodynamics, by Lenz
law,  ${\cal {\vec M}}$ is opposite to ${\vec B}$ (spin effects are
neglected), then ${\cal {\vec M}} \cdot {\vec B}<0$, and $P_{L\perp}>0$.
The opposite case occurs when ${\cal {\vec M}}$ is parallel to ${\vec
B}$, which occurs in the quantum case, i.e., when spin effects are
taken into account. We alert, however, that in the definitions and
derivations that follows use will be done of the magnitudes $|\vec B|$
and ${\cal{|\vec M|}}\equiv {\cal M}$ of both vectors $\vec B$ and
${\cal{\vec M}}$ instead of the vectors themselves.

\section{The energy-momentum tensor of a neutron gas}

Based on more fundamental grounds, one may write the general
structure of the energy-momentum tensor of a neutron gas in an
external field $\vec B$ in the same way as one does to construct
general tensors, as the polarization operator tensors
\cite{Shabad}, for instance. In an external field $F_{\mu \nu}$,
in addition to the basic  4-velocity vectors of the medium,
$u_\mu$, and particle momentum $k_\mu$, we have two extra vectors
$F_{\mu \nu}k_\nu, F^2_{\mu \nu}k_\nu$, to form a basis of
independent vectors (in what follows we shall use the notation
$F^2_{\mu \nu}=F_{\mu \lambda}F_{\lambda \nu})$ . From them we may
 build a set of basic tensors which, together with the tensors
 $\delta_{\mu \nu}$, $F_{\mu \nu}$, $F^2_{\mu \nu}$,
serves as a basis in terms of which we can expand any tensor
structure related to the particle dynamics, in particular, the
energy-momentum tensor. However, to get rid of tensor structures
containing off-diagonal terms, which would correspond to unwanted
shearing stresses in the fluid rest frame $u=(0,0,0,u_4)$, we
exclude some of them, i. e., $k_{\mu} k_{\nu}$, $k_\nu F_{\mu
\lambda}k_{\lambda}$, $F_{\mu \nu}$, $u_{\mu} k_{\nu}$, or any of
its combinations. By following the arguments used in
\cite{Shabad}, we conclude that we are left in the present case
with three basic tensors: $\delta_{\mu \nu}, F^2_{\mu \nu}, u_\mu
u_\nu$ to describe the dynamics of such a neutron gas.  Thus, the
structure of such an energy-momentum tensor is then expected to be
of the form

\begin{equation}
{\cal T}_{\mu \nu}= a\delta_{\mu
\nu}+ b F^2_{\mu \nu} + c u_\mu u_\nu
\label{EMTS}
\end{equation}

\noindent  where $\mu, \nu = 0, 1, 2, 3$; and $a = p$ is the
isotropic pressure term, $b={\cal M}/B$ and $c= U+p$. In the
present case the second of these tensors can be written in a
simpler form as $F^2_{\mu \nu}= -B^2\delta_{\mu \nu}^{\perp}$. The
tensor (\ref{EMTS}) has then the spatial eigenvalues $p-B{\cal M},
p-B{\cal M}, p$, and the time eigenvalue $-c=-U-p$, since $u_\mu
u_\mu=-1$. These eigenvalues exhibit the anisotropy in pressures
perpendicular and parallel to $B$.

In dealing with a quantum gas in an external field we shall assume that
the  sources of the field $\vec H$ are  either classical (currents) or
due to quantum effects. Although it is out of the scope of this paper
to discuss the mechanism for producing such field, we suggest, however,
some viable sources able to induce a self-consistent field, as for
instance a condensate of the vector meson $\rho $, the neutron
spin-spin ferromagnetic \cite{haensel-bonazzola96} coupling (see
below), or even by diquark \cite{horvath98} condensation.

 As in the case of the electron in an atom the
basic  dynamics in our present case is described by the Dirac
equation in an external field, in place of the Lorentz force.
According to the Ehrenfest theorem, classical dynamics is
contained on the adequate average of quantum dynamics. But quantum
dynamics leads also to several new phenomena not having classical
partner. After solving the Dirac equation one gets the energy
eigenvalue spectrum \cite{Ternov}. This energy depends on the
microscopic magnetic field $B$ through some interaction term in
the initial Lagrangian. These energy eigenvalues, after
the quantum statistical average, determine the thermodynamic
properties, such as the neutron gas pressure. If the coupling
constant is turned to zero, the particles would not feel any
pressure coming from the external field. Thus, since the classical
Maxwell stress tensor of the field $H$ does not express
by itself the  interaction with the particles, i. e., it expresses
the momentum and energy of the external field, we have no need to
add it to the expression (\ref{TPS}) below. The tensor (\ref{TPS})
 contains already the basic tensor structures of the
problem, including the Maxwell stress tensor of the field $B$,
which depends on the external field $H$ and the magnetization
${\cal{|\vec M|}}$.

The total external field $B$ contributes with virtual particles,
expressed through  the Euler-Heisenberg vacuum terms arising in
the regularization of the quantum vacuum terms \cite{Euler}. These
vacuum terms appear in the calculation of the basic statistical
quantity, the thermodynamical potential: $\Omega \equiv
-\beta^{-1}\ln{\cal Z}$, where $\cal Z$, the grand partition
function, is built up on the particle spectrum. Our
thermodynamical potential is the sum of two terms, $\Omega
=\Omega_{st}+\Omega_V$, the finite statistical term $\Omega_{st}$
plus the vacuum field contribution $\Omega_V$, which is divergent.
In the process of regularization, it absorbs the classical field
energy density term $B^2/8 \pi$. We observe here that Landau, in
p.69 of Ref.\cite{Landau0}, uses the specific term
\textit{thermodynamic potential}, denoted by $\Phi$ in that
reference, to name what
in western literature is known as Gibbs free energy, which is
denoted by G. Our thermodynamic potential is just what Landau
defines as ``new thermodynamic potential $\Omega$", but we have
taken it per
unit volume, which is dependent on $T$ and $\mu$ in absence of
external field, and in our case is dependent also on $B$. Observe
that $\Omega=F-G$. In the zero field case, it would be
$\Omega=-P$, where $P$ is the isotropic pressure. Due to the
spatial anisotropy introduced by the magnetic field $B$, the
pressures are not the same in all directions, and only in the
direction parallel to $B$ it adquires the value $\Omega (B)=-P_3$
(See below).

The coupling of the spin dipole moment of neutrons in an external
magnetic field $B$ produces a loss of rotational symmetry of the
particle spectrum (in what follows we will consider $B$ along the
$x_3$ axis). From the spectrum, which is expressed in terms of $B$,
by the standard methods of finite temperature quantum field
theory, we obtain the thermodynamical potential (per unit volume)
$\Omega =\Omega (B)$, and from it all the thermodynamic properties
of the system, in particular its magnetization, as is done by
Huang \cite{Huang} (p. 237), which is the statistical average
${\cal M} =-(\partial \Omega/\partial B)$.

Notice that this definition is consistent with our convention by
which thermodynamical quantities are defined in terms of the
microscopic magnetic field  $B$ acting on the particles as an
independent variable (see for instance Ref.\cite{Jackson}),
instead of using the quantity $H$. Thus, the thermodynamical
variable conjugated to the magnetic field $B$ is the quantity
${\cal M}(B)$, the system magnetization as introduced above. We
may write then

\begin{equation}
\Omega =-\int {\cal M}dB-P_0,  \label{Omega}
\end{equation}
\noindent

where $P_0 =- \Omega (0)$ is the term corresponding to the zero
magnetic field pressure. One must emphasize that in the quantum
relativistic case $\Omega$ depends on $B$ nonlinearly. From the
explicit expression for $\Omega$ given in another  section below,
one finds that the dependence of the energy spectrum on the particle
momentum
is not rotational invariant. This fact determines a reduction of
the symmetry  of the otherwise isotropic thermodynamic properties
of the system such as the pressure, which is expected to be
axially-symmetric for the reasons pointed out above.

By using the Green functions method it is found that the
energy-momentum tensor of matter in an external constant magnetic
field obeys the general structure (\ref{EMTS}),  and we have
\cite{Chaichian}

\begin{eqnarray}
{\cal T}_{\mu \nu }&=&(T\partial \Omega /\partial T+\sum \mu
_r\partial \Omega
/\partial \mu _r)\delta_{4\mu}\delta_{4\nu} \label{TPS} \\
&+&  4F_{\mu \lambda }F_{\nu \lambda }\partial
\Omega /\partial F^2 - \delta _{\mu \nu
}\Omega,\nonumber  \end{eqnarray}

\noindent where $r$ run over the species involved. Below we will
take $r=n,p,e$ to describe teh neutron, proton and electron
component of the star gas. Expression (\ref{TPS}) in the zero
field limit
reproduces the usual isotropic energy-momentum tensor ${\cal
T}_{\mu \nu }=P\delta _{\mu \nu }-(P+U)\delta _{4\mu }\delta _{\nu
4}$ of a perfect fluid. From Eq.(\ref{TPS}) the spatial components
are ${\cal T}_{33}=P_3 =-\Omega ,{\cal T}_{11}={\cal
T}_{22}=P_{\perp }= - \Omega -B{\cal M}$. The time component
${\cal T}_{00} \equiv -U_r= -TS_r- \mu_r N_r-\Omega_r$, in which
$U_r \sim \mu_r N_r \sim \Omega_r$ are quantities of the same
order of magnitude. We shall assume that inside a stable NS
there is locally a
balance between the mechanical stresses (and forces) from the
pressure exerted by the quantum gas and the gravitational stress,
i. e., the gravitational force per unit area, exerted by the star
mass, as defined below by equations (\ref{PG}),(\ref{PG1}). By
abusing a bit of the phrasing we shall refer over the paper to
this last one as the ``gravitational pressure", in an attempt to
turn more meaningful discussions regarding effects of the gas
pressure and those stemming from its own gravity. In the case of a
neutron star the correct treatment of this problem requires of
general relativistic corrections\cite{Weinberg}. However, for the
general purpose of this paper we consider here the problem in a
rough manner, in a similar fashion as it was done in the white
dwarf case with zero magnetic field \cite{Huang}, in Newtonian
gravity. If $dS_i$ is an element of surface and $dx_i$ an increase
in the coordinate $i=1,2,3$, one can express the balance between
the work done on the star and the variation of its gravitational
energy, $E_g \equiv G M^2 R^{-1}$, under a small change of volume
$\Delta V$ as

\begin{equation}
\int(- {\cal T}_{ij}dS_j  +  \frac{\partial E_g}{\partial
x_{i}})dx_i=0 \; ,
\end{equation}

\noindent


where the quantity in parenthesis must be zero, and describes the
balance of forces. We will use now cylindrical coordinates with the $z$
axis parallel to $B$, and assume that the shape of the deformed star
can be (very roughly) approximated by the simplified model of a cylinder
of height $Z$ and radius
$r_{\perp}$ (we could equally use as a simple model a cylinder of
height $Z$ ended by two half-spheres of radius $r_{\perp}$. The
equations (\ref{PG}) and (\ref{PG1}) below would differ by  unimportant
numerical factors). We assume also that the change of the cylinder's
volume $\Delta V$ and surface $\Delta S$ are small, so that the total
surface of the star $S= 2\pi r_{\perp}(r_{\perp}+ Z)$ is approximately
constant. Then we have $dS_{\perp}=2\pi r_{\perp} dz$, $dS_{3}=2\pi
r_{\perp} dr_{\perp}$ and integrating we get the equilibrium between
the gravitational and gas pressures $P_{\perp}=P_{g\perp}$ and
$P_{33}=P_{g33} $, where

\begin{equation}
P_{g\perp}=\frac{1}{2\pi r_{\perp}Z}\frac{\partial E_g}{\partial
r_{\perp}}. \label{PG}
\end{equation}
\noindent and
\begin{equation}
P_{g33}=\frac{1}{\pi r_{\perp}^2}\frac{\partial E_g}{\partial z}.
\label{PG1}
\end{equation}

Eqs.(\ref{PG}),(\ref{PG1}) are to be interpreted
respectively as the transverse and longitudinal "gravitational
pressures" in the sense defined above. By assuming that
$\frac{\partial E_g}{\partial r_{\perp}}$ and $\frac{\partial
E_g}{\partial z}$ are quantities of the same order, to preserve the
balance of transverse pressures when $P_{\perp}$ decreases (due to the
increase of the product $B{\cal M}$), the quantity $1/2\pi r_{\perp}Z$
must decrease. This is achieved whenever $Z$ increases faster than
$r_{\perp}$ decreases (all this can be seen by assuming that the total
surface $S=const$ and $ dr_{\perp}/d\tau<0$, where $\tau$ is a
parameter like time). In that case the outcome is a stretching of the
body along the direction of $B$. Thus, the anisotropy of pressures in
our problem suggests that matter in the body is also distributed
anisotropically, what leads to prolate isobaric surfaces \footnote{
This is an effect opposite to the oblateness of the Sun, Earth and
other planets due to the effective decrease of the transverse
gravitational force by the centrifugal force.}.

The condition $P_{\perp}=0$ implies also that $U \sim B {\cal M}$.
As $P_3> 0$, an instability arises in the system leading to a
transverse collapse. Thus, as pointed out in the introduction, new
physics arise in connection with the vanishing of the transverse
pressure, $P_{\perp}=0$. This peculiar behavior will be discussed
in the forthcoming sections.

By writing ${\cal M}=(B-H)/4\pi $, one may  formally write $\Omega
=-\frac{1}{8\pi }B^2+\frac{ 1}{4\pi }\int HdB-P_0$. We remind that
as $\Omega \equiv F-G$, the last expression is consistent with
what would be obtained in the classical nonrelativistic case
\cite{Landau}, where $F=F_0+\int HdB/4\pi $ is the Helmholtz free
energy and $G=F+\int {\cal M}dB+P_0$ $=G_0+B^2/8\pi $ as the Gibbs
free energy. Because of our definition of ${\cal M}$, the last
term is given in terms of $B$ and not in terms of $H$. As we have

\begin{equation}
{\cal T}_{33}=-\Omega=\frac{1}{8\pi }B^2-\frac{ 1}{4\pi }\int HdB+P_0,
\end{equation}

and also

\begin{equation}
{\cal T}_{11}={\cal T}_{22}=-\Omega -B{\cal M}=-\frac{ 1} {8\pi
}B^2+\frac{
1}{4\pi }\int B dH+P_0,
\end{equation}

it is straightforward to see that the spatial components of the
energy-momentum tensor ${\cal T}_{ij}$ ($i,j=1,2,3$) can be
rewritten as

\begin{equation}
{\cal T}_{ij}= P_0\delta _{i j} -{\cal T}^M_{i j} (B, H)+ {\cal
S}(B)_{i j}, \label{MMIN}
\end{equation}

where ${\cal S}(B)_{i j}=\frac{1}{4\pi}[B_i B_j
-\frac{1}{2}(B^2)\delta_{i j}]$ is the Maxwell stress tensor for
the microscopic field $B$, and ${\cal T}^M_{i j}
(B)=\frac{1}{4\pi}[H_i B_j -(\int B dH)\delta_{i j}]$ is the
Minkowski tensor for relativistic nonlinear media  (see below).
This last term  reduces to the usual expression in the nonrelativistic
limit if $H$ depends linearly on $B$ \cite{Jackson}. The
definition (\ref{MMIN}) expresses the total pressure as a sum of
an isotropic pure mechanical pressure (independent of the
electromagnetic field) plus a pressure coming from the Minkowski
tensor due to the interaction of the external field $H$ with the
microscopic field $B$, plus the Maxwell stress tensor of the
microscopic field $B$. If $H=0$, $B=4\pi{\cal M}$ and ${\cal
T}_{ij}=P_0\delta _{i j} + {\cal S}(B)_{i j}$. In this case the
magnetic field is kept self-consistently.

At this point we want to refer to the recent paper by Khalilov
\cite{Khalilov} where a similar problem to the present one is
studied. However, the expression for the stress tensor in a medium
is taken as given only by the linear approximation of the
Minkowski tensor term. This approach is not justified in the
relativistic case. Further, the Maxwell tensor of the microscopic
field $B$ and the isotropic $P_0\delta_{ij}$ terms are omitted.
This leads the author to wrongly conclude that the collapse occurs
like in the classical case (see below), that is, along the $\bf B$
field, in contradiction to our present results and those of
\cite{Hugo}, \cite{Chaichian}. A consistent approach, as we have
followed here, and discussed in the accompanying paper
\cite{nos04}, leads to opposite results compared to those of Ref.
\cite{Khalilov} in what concerns the spatial direction  of
collapse, although the fundamental issue regarding the collapse of
the neutron star remains taking place in both theories.

Note, in addition, that  Refs. \cite{Herrera} and \cite{Mak},
where  the problem was also studied, did not take into account the
dynamical effects of the strong (and super strong) magnetic fields
supposed to exist in the core of canonical NSs, see for instance
Refs. \cite{Chaichian},\cite{Cardall}, \cite{Chakrabarty},
\cite{Broderick}. Thence, it is the contention of this paper to
address this open issue. The novel results obtained here point out
to the occurrence of new physics and processes in the
(relativistic) astrophysics of compact objects that were not
manifest in previous papers.

\section{ Classical vs. Quantum Collapses}

In Ref.\cite{Chaichian} we found that a relativistic degenerate
electron gas placed in a strong external magnetic field $B$ is
confined to a finite set of Landau quantum states. As the field is
increased the maximum Landau quantum number is decreased favoring
the arising of either a paramagnetic or a ferromagnetic response
through a
positive magnetization ${\cal M}$,  up to the case in which only
the ground state is occupied. The gas then becomes topologically
one dimensional, and in consequence the pressure transverse to the
field vanishes for fields $B = \Omega/{\cal M}$ \cite{Chaichian}.
Thus, the electron gas becomes unstable due to the decrease of the
transverse pressure for fields strong enough, and the outcome is a
collapse.

For neutrons the magnetization is always positive (see arguments
below) and nonlinear, what leads to a sort of ferromagnetic
behavior. For fields strong enough the  pressure transverse to the
field, $P_{\perp }=-\Omega -B{\cal M}$, is considerably decreased
and may vanish. If we assume that extremely magnetized NSs, as the
Duncan and Thompson magnetars\cite{DT92}, have fields $H\sim
10^{15}$ G, and that
inside the star $B$ increases by following a dipole law
$B(r^{\prime})=B_{surf}/r^3$, we expect near its surface magnetic
fields ranging from $10^{16}-10^{17}$G\cite{DT92,Kondratyev02} up
to values of order $10^{20}$G in its core\cite{Chakrabarty},
where the field is
maintained self-consistently, i. e., $H=0$. For fields of this
order of magnitude super dense matter composed of neutral
particles having a magnetic moment may undergo a transverse
collapse since $P_{\perp }$ vanishes. As discussed below, the
emerging physics seems to ban the possibility of magnetar
formation. The outcome of such a collapse might be a compact star
endowed with canonical magnetic field, as discussed below.

In the classical case in which the response of the medium is due
to the Lenz law, the magnetization is opposite to the external
field $H$ and it may happen that ${\cal M}<0$. This also occurs in
the diamagnetic case. Then $H>B$ and $ P_{\perp }>P_3$. Note that
the opposite occurs in some permeable materials where ${\cal M}>0$
and $H=B-4\pi{\cal M}$ is small in comparison to either ${\cal M}$
and $B$; this is due to ferromagnetic effects which have quantum
origin, as in the neutron gas.

In the case of a classical magnetized gas, as $ P_{\perp}>P_3$, this
leads to the Earth-like oblatening effect described above. But
opposite to this, for the
critical quantum configuration of the NS gas the coupling of the
spin magnetic dipole with the magnetic field ${\vec B }$ plays the
main role, and ${\cal M}>0$ (see Eq.(\ref{test}) below and the
subsequent discussion where this is shown explicitly), leading to
ferromagnetic effects. The situation then is reversed and
$P_{\perp }$ is smaller than $P_3$ in the amount $B{\cal M}$ and
it vanishes for $\frac 1{ 8\pi }B^2=\frac 1{4\pi }\int BdH+P_0$
leading, conversely, to a prolate configuration.

In classical electrodynamics \cite{Landau3} it is suggested that
the total pressure is given by the sum of the Maxwell stress
tensor ${\cal S}_{\mu \nu }$ plus an isotropic pressure ($P_0$)
term. In the case of a constant magnetic field parallel to the
$x_3$ axis, the total pressure tensor reads ${\cal
T}_{ij}=P_0\delta _{ij}+{\cal S}_{ij}$
 or $P_3= P_0-B^2/8\pi$ and $P_{\perp }=P_0+B^2/8\pi.$

As pointed out before, the body deforms under the action of these
anisotropic pressures. If the longitudinal pressure decreases, the
body flattens along the magnetic field \cite{Konno}. Thus, in this
pure classical case, for the extreme limit of flattening, $P_3=0$
and $P_{\perp}=P_0+ B^2/8\pi $, and the body would collapse as a
disk or a ring perpendicular to the field. Starting from general
relativistic considerations it has been reported \cite{Cardall}
the existence of a maximum magnetic field for having stationary
configurations of NSs. (We interprete that result as indicating the
occurrence of a classical collapse for fields larger than the
maximum). This field induces a toroidal
configuration, which is topologically equivalent to a ring. In the
quantum case, for degenerate fermions, as ${\cal M}>0,$ it is
$P_{\perp }=-\Omega -B{\cal M}$ which is decreased by increasing
$B$. As the NS is in equilibrium under the balance of neutron and
"gravitational" pressures, the body stretches along the direction
of the magnetic field. Thus, for any density there are values of
the field $B$ high enough such that these pressures cannot
counterbalance each other leading to a collapse perpendicular to
the field for $P_{\perp }=0$. This collapse would leave as a
remnant a nucleons plus a Bose-Einstein-like condensate,
a hybrid or strange star \footnote{The reader is addressed to the
interesting references: "What if pulsars are born as strange
stars?", by R. Xu, B. Zhang
and G. Qiao, Astropart. Phys. 15, 101 (2001), and "PSR 0943+10:
A bare strange star?", by the same authors and published in Astrophys.
J. 552, L109 (1999).} with canonical magnetic field
\cite{horvath99}, or a distorted ("cigar-like") black hole.

Our previous considerations are approximation-independent. In
order to discuss an specific model, we shall start by computing
the free particle spectrum for neutral particles in a magnetic
field.

\section{The Neutron Gas in a Magnetic Field}

For free neutrons in a magnetic field $B$ we have the Dirac
equation for neutral particles with anomalous magnetic moment
\cite{Ternov}

\begin{equation}
 (\gamma_{\mu}\partial_{\mu} + m +iq\sigma_{\mu
\lambda}F_{\lambda \mu })\psi=0, \end{equation}

 where $\sigma_{\mu \lambda}=\frac{1}{2}(\gamma_\mu \gamma_\lambda-
 \gamma_\lambda \gamma_\mu)$ is the spin tensor, and $F_{\lambda \mu }$
 is the electromagnetic field tensor describing $B$. By solving this
equation we get the eigenvalues \cite{Ternov}

\begin{equation}
E\hspace{0in}_n(p,B,\eta )=\sqrt{p_3^2+(\sqrt{p_{\perp }^2+m_n^2}+\eta
qB)^2},
\label{ein}
\end{equation}

\noindent where $p_3$, $p_{\perp }$ are respectively the momentum
components along and perpendicular to the magnetic field $B$, $m_n$ is
the neutron mass, $q=1.91M_n$ ($M_n$ is the nuclear magneton), $\eta
=1,-1$ are the $\sigma _3$ eigenvalues corresponding to the two
orientations of the magnetic moment (parallel and antiparallel) with
regards to the field $B$. Here we make the assumption that the magnetic
moment remains constant for large magnetic fields. Actually, this is
not so.  Radiative corrections change its value as a function of $B$.
This problem is, however, beyond the scope of the present approach.

From (\ref{ein}) we see that although the Hamiltonian is linear
in $B$, the eigenvalues depend on $B$
nonlinearly. This makes the relativistic thermodynamic and
electromagnetic properties of the system of neutrons very
different from the nonrelativistic case. For instance, all
thermodynamic quantities, $\Omega$, ${\cal M}$ (and in
consequence, $H=B-4\pi{\cal M}$), and $N$ are also nonlinear
functions of $B$. The expression (\ref{ein}) also shows manifestly
the change of spherical to axial symmetry with regard to momentum
components. {\it This anisotropy in the dynamics is expected to be
reflected in an anisotropy in the thermodynamic properties of the
system, as it is expressed by the difference between the
transverse and longitudinal pressures discussed earlier on an
approximation-independent basis.}

The partition function ${\cal Z}={\rm Tr}(\rho )$ is obtained from
the density matrix describing the model $\rho =e^{-\beta \int
d^3x({\cal H}({\bf x)}-\sum \mu _r N_r)}$ \cite{Hugo1}. Here $\mu
_j$ ($j=1,2,3$) are the chemical potentials associated with the lepton,
baryon and electric charge conservation, so that $\mu _n=\mu _2$,
$\mu_p=\mu _2+\mu _3,\mu _e=\mu _1+\mu _3$ and $\mu _\upsilon =\mu
_1$. The thermodynamical potential can be written as $\Omega
=-\beta ^{-1}ln{\cal Z}$. Usually the eigenvalues of ${\cal H}$
contain the contribution from neutrons, protons, electrons and
some meson species, and the densities are $N_r=-\partial \Omega
/\partial \mu _r$, where $r=n,p,e...$ We name $\Omega
=\sum_r\Omega _r $, ${\cal M=}\sum_r{\cal M}_r$ the total
thermodynamical potential and magnetization, respectively.

A standard procedure is to work in the mean field approximation in
which the meson fields $\sigma ,\rho ,\omega $ are taken as
constant, as done in Refs.\cite{Chakrabarty}, \cite{Broderick}, through
which the mass spectrum of baryons is corrected and strong
repulsive interactions between them is found. However, for
simplicity we will keep the spectra in the tree approximation to
obtain the one-loop approximation for $\Omega $, and neglect the
statistical contribution from meson terms in $\Omega $ as compared
with those of fermions (for them $m_i\beta \sim 10^3-10^4$)
except for fields $B\lesssim B_{c\rho }=m_\rho ^2/e\approx
10^{20}$G, since the contribution of the $\rho $ vector meson
condensate to ${\cal M}$ becomes relevant, and in analogy with
$W^{\pm }$-s\cite{Chaichian}, leads to a self-consistent
spontaneous magnetization $B=4\pi {\cal M}=2 \pi eN_\rho
\sqrt{m_\rho ^2-eB}$, where $N_\rho$ is the condensate density.
However, for such fields the magnetic pressure: $\int {\cal M}dB-{\cal
M}B=-B^2/8\pi $ overwhelms the kinetic pressure term $P_0$ (of
order $10^{36}$ dynes/cm$^2$) leading to $P_{\perp }<0,$ and the
star is definitely unstable: it collapses. This mechanism is valid
for other quasi-particle vector boson condensates, as di-quarks,
which may be formed in the medium even for smaller values of the
field $B$ \cite{horvath99,horvath98}.

\section{Thermodynamical potential of a neutron gas}

The Green functions method \cite{Fradkin} leads to a general
expression for the relativistic thermodynamical potential. At the
one-loop level, where no radiative corrections are considered, it
is a generalization of the usual nonrelativistic formula because
of the fact that antiparticles must also be included. Particles
and antiparticles contribute with chemical potentials of opposite
sign, leading to sums or integrals over the quantum numbers
involved, of terms containing the product of the logarithms of
$(e^{-(\mu_n \pm E_n )}+1)$. In this case, when no external fields
are present, a divergent term accounting for the vacuum energy
appears which is usually subtracted \cite{Fradkin}. In presence of
an external field, however, a term accounting for the vacuum
contribution, must also be included: The Euler-Heisenberg energy
of vacuum in an external field \cite{Shabad,Euler,Hugo1,Weisskopf}.
One can obtain an expression for the thermodynamic potential of the
neutron gas in the one-loop approximation as $\Omega_n=
\Omega_{sn} + \Omega_{Vn}$, with

\begin{eqnarray}
\Omega_{sn}& =& -\frac{1}{4 \pi^2 \beta}\sum_{\eta=1,-1}
 \int_{0}^{\infty}p_{\perp} dp_{\perp} dp_3 \ln \left[f^+(\mu_n, \beta)
f^-(\mu_n, \beta) \right], \label{ONT}
\end{eqnarray}

where $f^{\pm}(\mu_n, \beta)=(1 + e^{-(E_n \mp \mu_n)\beta})$
accounts, respectively, for the contribution of particles and
antiparticles.  The expression for the vacuum term reads thus

\begin{eqnarray}
\Omega_{Vn}& =& \frac{1}{4 \pi^2 \beta}\sum_{\eta=1,-1}
 \int_{0}^{\infty}p_{\perp} dp_{\perp} dp_3 E_n \label{ONT1}
\end{eqnarray}

\noindent which is divergent. In the Appendix we will show how to
regularize this expression, and how to obtain the analog to the
Euler-Heisenberg energy of vacuum due to the neutrons
contribution.

 After integrating by parts in (\ref{ONT}) its
evaluation becomes easier. The Fermi distributions, which arise
by differentiating $f^{\pm}$ with respect to $p_3$, are
$n^{\pm}=1/(1 + e^{(E_n \mp \mu_n)\beta})$. In the degenerate case
this expression reduces in $n^-= \theta (\mu - E_n)$ and $n^{+}=0$,
since in that case only particles contribute to $\Omega$. The
resulting expression splits itself in two terms where the
integrals are bounded by the Fermi surfaces $\mu - E_n (\eta=\pm
1)=0$. These surfaces have axial symmetry, and thus, the Fermi
momentum is not a definite number, given only in terms of $\mu_n$
and $m_n$, but on the opposite, it has infinite values. Thence, we
have

\begin{eqnarray}
\Omega_{sn}& =& -\frac{1}{4 \pi^2}\sum_{\eta=1,-1}
 \int_{0}^{\infty}p_{\perp} dp_{\perp} \frac{p_3 dp_3}{E_n}\theta
\left(\mu - E_n (\eta)\right) \; . \label{ONT1}
\end{eqnarray}

The $\theta$ function  bound
 these integrals in the intervals
$-p_{3F} \leq p_3 \leq p_{3F}$, where
$p_{3F}=\sqrt{\mu^2-(\sqrt{p_{\perp}^2+m_n^2}+\eta y)}$ and $0
\leq p_{\perp}\leq \sqrt{(\mu-\eta y)^2-m_n^2 }$. After some
transformations, it yields

\begin{eqnarray}
\Omega_{sn} &=&-\Omega _0\sum_{\eta
=1,-1}\left[\frac{xf^3}{12}+\frac{(1+\eta y)(5\eta
y-3)xf}{24}\right. \nonumber  \\ & + & \left. \frac{(1+\eta
y)^3(3-\eta y)}{24}L-\frac{\eta yx^3}6s\right],
 \label{omega}
\end{eqnarray}
\noindent where $x=\mu _n/m_n,$ ($m_n(x-1)$ is the usual Fermi
energy), and $y=qB/m.$ We define the functions $f\equiv f(x,\eta
y)=\sqrt{x^2-(1+\eta y)^2}$, $s\equiv s(x,\eta y)=(\pi /2-\sin
^{-1}(1+\eta y)/x)$, $L\equiv L(x,\eta y)=\ln (x+f(x,\eta
y))/(1+\eta y)$. The functions $f = <p_F>/m_n$, where $<p_F> =
\sqrt{\mu_n^2-(m_n +\eta qB)^2}$ and we name them the average
Fermi momenta for $\eta=\pm 1$. We see that $m_n-\eta qB $ behaves
formally as a two-valued magnetic mass.

In the zero field limit one gets $P_0=-\Omega_n (y=0)$, where
\begin{eqnarray}
\Omega_{sn} (y=0) &=&-\Omega
_0\left[\frac{xf_0^3}{12}-\frac{xf_0}{8}+ \frac{1}{8}L_0\right],
 \label{omega}
\end{eqnarray}
where $f_0=\sqrt{x^2-1}$ is the  relative Fermi momentum $p_F/m$,
and $L_0 =\ln (x+f_0)$.
 The  neutron vacuum term (see Appendix) has an
Euler-Heisenberg-like form as

\begin{eqnarray}
\Omega_{Vn}&=&\frac{1}{4\pi^2}\int_0^{\infty}dy
y^{-3} e^{-(m_n^2+q^2B^2) y} [\cosh(qBmy)-1-(qBmy)^2/2!] \label{nvac}\\
\nonumber &+& \frac{q B}{2\pi^2}\int_0^{\infty}dy
y^{-2}\int_0^\infty dw e^{-[(w+m_n)^2+ q^2
B^2]y}[\sinh(2qB(w+m_n)y)-(2qB(w+m_n)y)+
(2qB(w+m_n)y)^3/3!]
\end{eqnarray}

\noindent It can be shown (see Appendix) that the more significant
term in (\ref{nvac}) is the first one, which for fields of order
$10^{17}$ G leads to $\Omega_{Vn} \sim 10^{30}$
erg$\cdot$cm$^{-3}$ and is negligible small as compared with
$\Omega_{sn}$ up to $B \sim 10^{18}$ G. Thus, we neglect it in a
first approximation in what follows. We must point out, however,
that since neutrons have a quark structure, a more fundamental
quantity would be the vacuum quark contribution, whose order of
magnitude is expected to be near $\Omega_{Vn}$. Apart from this
note, it should be emphasized that the role of vacuum cannot be
ignored for fields greater than $10^{18}$ G.

From $N_n =\partial\Omega/\partial \mu _n$ one gets

\begin{eqnarray}
N_n &=&N_0\sum_{\eta =1,-1}\left[\frac{f^3}3+\frac{\eta y(1+\eta
y)f)}2-
\frac{\eta yx^2}2s\right], \label{den}
\end{eqnarray}

In the limit $B=0$ Eq.(\ref{den}) reproduces the usual density  of
a relativistic Fermi gas at zero temperature, $N_n = N_0 f_0^3/3$

Having an equation relating the chemical potentials, and demanding
conservation of both baryonic number $N_n+N_p=N_B$ and electric
charge $ N_p+N_e=0$, in principle one may solve exactly the
problem in terms of the external field as a parameter.
Nonetheless, we shall focus our discussion on the properties of
the equation of state. Note in passing that our expressions for
the spectra and densities of neutrons and protons are similar to
those of Ref. \cite{Mathews} of a neutron gas in a magnetic field,
but we get different equations of state.

Finally, for the magnetization, given as ${\cal M}_n=- \partial
\Omega_n/\partial B$, we have

\begin{eqnarray}
{\cal M}_n & = & -{\cal M}_0\sum_{\eta =1,-1}\eta \left[\frac{(1-2\eta
y)xf}6
\right. \\
& - & \left. \frac{(1+\eta y)^2(1-\eta y/2)} 3L + \frac{x^3}6s\right]
\nonumber
\end{eqnarray}

\noindent

where $N_0=m_n^3/4\pi ^2\sim 2.04\times 10^{39}$, $\Omega
_0=N_0m_n\sim 3.0\times 10^{36}$, and ${\cal M}_0=N_0q\sim
2.92\times 10^{16}$ and one can write ${\cal M}_n={\cal
M}_n^{+}(\eta =-1)-{\cal M}_n^{+}(\eta =+1)$, and obviously,
${\cal M}_n\geq 0$.

We confirmed by explicit calculation that ${\cal M}$ is a nonlinear
function of $B$ and, in this sense,
the magnetic response is ferromagnetic. A fully ferromagnetic
response demands to include also the spin-spin coupling
contribution. We discuss briefly this point below.

To see why the magnetization is always positive for the neutron
gas note that the magnetic susceptibility $\chi=\partial {\cal
M}_n/\partial B$ can be easily obtained as

\begin{equation}
\chi= \frac{q {\cal M}_0}{2m_n}\sum_{\eta =\pm 1} [xf+ (1+\eta
y)^2 L],\label{test}
\end{equation}

\noindent which for $x>1$, $y \leq 1$, and $f$ real and positive,
it is $\chi>0$. This means that ${\cal M}_n$ is an increasing
function of $B$ (or $y$) under such conditions. As ${\cal M}_n
(y=0)=0$ and ${\cal M}_n (y=1)= 2{\cal M}_0 (1+\pi)>0$, this means
that ${\cal M}_n >0 $ in the region between these two points,
which is the one of interest for us (the region I discussed
below and showed in Fig. 2).

The fact that we are summing over the magnetic moments oriented
parallel ($\eta=-1$) and antiparallel ($\eta=+1$) to $B$ is
similar to the well known Pauli paramagnetism in nonrelativistic
quantum statistics. We may consider each term $\eta=\pm 1$ as
representing a phase of the system.   One can draw in the $x,y$
plane, for both $x>0$, $y>0$, two regions limited by the lines
$x-y = \pm 1$ and $x+y =1$ (see Figure 1). The region I is bounded
above by the line $x=y+1$, on which the contribution from $f(\eta
=+1)$, and all other terms containing $\eta=+1$ vanish: all
magnetic moments are aligned parallel to $B$. Below, such region
is limited by $y=0$. For points in the region $x>y+1$ and $y>0$
both terms containing $\eta=-1, +1$ are nonzero and real. Thus,
the quantities $\Omega, N_n$, and ${\cal M}$ are real and although
 most neutrons have their magnetic moments along $B$, there are
some amount of them having their magnetic moments antiparallel
to $B$. In this region, near the line $x=y+1$, is located the
curve $P_{\perp}=0$, which we will discuss below.

In the region II  limited by $x \geq 1-y$, $y+1
> x \geq y-1$ only the term $\eta=-1$ is real and the term
$\eta=+1$ becomes purely imaginary, and all neutrons would have
their magnetic moments parallel to $B$. The term $\eta=-1$
contribute to $\Omega, N_n$, and ${\cal M}$. For $x<y-1$, there
are no physical solutions. For $y=0$, the magnetic field, and in
consequence, the magnetization ${\cal M}$ vanish. For $y<0$
(magnetic field along the negative $y$ axis) we have the reverse
situation to the one described above (See Figure 1).

With regards to the background of electrons and protons, we remind
that the case of the electron gas was discussed in
(\cite{Chaichian}). We expect that for magnetic fields of order
$B_{ce} =m_e^2/e \sim 10^{13}$ G and densities around $10^{30}$
cm$^{-3}$ all electrons are in the Landau ground state, and the
system show the instability which arises from the vanishing of the
transverse pressure. For densities and magnetic fields above these
critical values, the stability of the electron gas is doubtful.
Recently Ghosh, Mandal and Chakrabarty \cite{Chakrabarty1} has
pointed out strong arguments against the equilibrium of an
electron gas under $\beta$ decay in strongly magnetized neutron
stars. The situation may be modified if radiative corrections are
taken into account: if $ B_{ce}=m_e^2/e\sim 10^{13}$G is the usual
QED critical field, for larger fields of order $ B_{ce}^{\prime
}\sim 4\pi /\alpha B_{ce}\sim 10^{16}$G the contribution of the
electron anomalous magnetic moment becomes significant and the
problem cannot be satisfactory treated at the tree level. However,
starting from the results of \cite{Chaichian},\cite{Chakrabarty1},
one concludes that the electron gas is hardly in equilibrium for
fields beyond $B_{ce}$. One possibility is  the bosonization of
the electron system, as has been recently suggested
\cite{Chakrabarty2}. This may be accomplished through the increase
of the spin-spin interaction, leading to the formation of parallel
spin electron pairs, equivalent to charged vector bosons.

If we include both the normal and the anomalous magnetic moments
for protons, one can give a formula for their spectrum in the
external field $B$ as\cite{Ternov}:

\be E_{p}=\sqrt{p_3^2+(\sqrt{2eBn+m_{p}^2}+\eta q_{p}B)^2}, \ee

 $ q_p=2.79M_n$. For neutrons, the critical field at which the
coupling energy of its magnetic moment equals the rest energy is
$B_{cn}=1.57\times 10^{20}$ G. For protons it is $B_{cp}=2.29\times
10^{20}$ G. By defining $x_p=\mu _p/m_p$, $y_p=q_p/m_p$,
$b=2e/m_p^2$, then $y_p=2.79e/2m_p^2$. We name also $ g\equiv
g(x_p,B,n)=\sqrt{x_p^2-h(B,n)^2}$ and $h\equiv
h(B,n)=(\sqrt{bBn+1} +\eta y_pB)$. Thus for the proton
thermodynamical potential we get

\begin{eqnarray}
\Omega _p &=&-\frac{eBm_p^2}{4\pi ^2} \sum_n\sum_{\pm \eta
}\left[x_pg-h^2\ln (x_p+g)/h\right],
\end{eqnarray}

and for its density

\be
N_p=\frac{eBm_p}{2\pi ^2}\sum_n\sum_{\pm
\eta}g(x_p,B,n)\;,
\ee

 while the magnetization is given by

\begin{eqnarray}
{\cal M}_p &=&\frac{em_p^2}{4\pi ^2}\sum_n\sum_{\pm \eta} \left\{x_pg -
\left[h^2+(\eta y_p \right. \right. \\
& + & \left. \left. (bn/2\sqrt{bBn+1}))\right] \times \ln\left(x_p +
g\right)/h\right\},  \nonumber
\end{eqnarray}

\noindent where the coefficients of these formulae are
$N_{0p}=em_pB/2\pi ^2\sim 4.06\times 10^{19}B$,
$\Omega_{0p}=N_{0p} m_p B\sim 6.1\times 10^{16}B$, and ${\cal
M}_{0p}=N_{0p}m_p=\Omega _0/B$. The maximum occupied Landau
quantum number $n$ may be given as $n_{max}=(x_p-\eta
y_pB)^2-1/bB$. For $B\ll B_{cp}$, so that $y_pB\ll 1$, and
$x_p\geq 1$, one can take approximately $n_{max}\sim
(x_p^2-1)/bB$, and for fields large enough $ n_{max}=0$. We expect
that from the equation $\mu_n=\mu_p+\mu_e$, then $x_p \sim x_n$,
the previous expression for the proton density $N_p$ decreases
with increasing $ B $, favoring the inverse beta decay. For fields
$B\sim m_p/q_p$ and $x_p\gg 1$, $n_{max}\geq 1$, and thus large
Landau numbers are again occupied. However, for $x_n , x_p \gtrsim
1$, being both quantities of the same order of magnitude, from the
comparison of $\Omega_{0p},N_{0p},{\cal M}_{0p}$ with
$\Omega_{0},N_{0}, {\cal M}_{0}$ we conclude that for fields below
$10^{19}$ G, the dominant longitudinal pressure, density and
magnetization comes from the neutron gas.

\begin{figure}
\begin{center}
\end{center}
\vskip 0.5truecm
\caption{This plot shows the regions in the $x$,$y$ plane
where the neutron magnetic moments are oriented
parallel or antiparallel to the magnetic field ${\vec B}$. Special
attention should be given to regions I and II, where the solutions
discussed in the text are valid. }
\end{figure}

\section{Condition for zero transverse pressure and collapse}

In the electron gas case \cite{Chaichian}, the
vanishing of the transverse momentum can be guessed from the spectrum
when all the system is confined to the Landau ground state. The
spectrum corresponds to a  purely one-dimensional system moving along the
external field, and the transverse Fermi momentum is zero. In a similar
way,in the neutron gas case we observe that the threshold of zero
transverse pressure $P_{\perp }=0$
can be figured out from the spectrum described by Eq.(\ref{ein}), since
the contribution from the $ \eta =-1$ term is dominant (observe that
the term with $\eta =+1$ contributes with a negative term to
${\cal M} $). We shall consider on the Fermi surface for $\eta=-1$
the quantity

\begin{equation}
p_{F\perp eff}^2 =\mu_n^2-p_{F3}^2-m_n^2 =\left(\sqrt{
p_{F\perp}^2 + m_n^2}-qB\right)^2-m_n^2
\end{equation}

\noindent which we name the effective squared Fermi transverse momentum.
If $B \ll 10^{20}$ G, then $q^2 B^2 \ll 2e B m_n$, the
vanishing of $p_{F\perp eff}$ is guaranteed if

\begin{equation}
p_{F\perp} \sim \sqrt{2q B m_n}.\label{heu}
\end{equation}

\noindent The resulting Fermi surface would be equivalent to that
for one-dimensional motion, parallel to $B$, i. e., for particles
having energy $E_n\simeq \sqrt{ p_3^2+m_n^2}$, and in consequence
the transverse momentum (and pressure)  vanishes. Notice
from (\ref{heu}) that for $p_{F\perp}/m_n \sim 10^{-1.5}$ one has
$qB/m_n=y \sim 10^{-3}$, which means fields of order $10^{17}$G. A
more accurate result is obtained, however, from the condition:
${\cal T}_{\perp }=0$.

In Figure 2 we have drawn the equation $P_{\perp n}=-\Omega
-B{\cal M}=0$ in terms of the variables $x,y$. We observe that
there is a continuous range of values of $x$, starting from
$x=1.005$ to $x=1.125$ and from $y=0.001$ to $ y=0.1$, for which
the collapse takes place. The latter values of $y$ means fields in
the interval $B\sim 10^{17}$ to $10^{19}$ G. To these ranges of $
x,y$ corresponds a continuum range of densities, from
$10^{-2}N_0$
($ 10^{12} $ g/cm$^{-3}$) onwards. The transverse compression of
the whole mass of the star due to flux conservation leads to an
increase of $B$ and the mechanism of collapse is enhanced.

\begin{figure}
\begin{center}
\end{center}
\caption{The
instability condition: $P_{\perp} = 0$ in terms of the variables $x$,
$y$. A nascent strongly magnetized neutron star having a configuration
such that its dynamical stage would be represented by a point above the
central curve in this plot would be unstable to transversal collapse,
since $P_{\perp} \leq 0$ there. }
\end{figure}

\subsection{The spin-spin coupling contribution}

Although we used previously the term ferromagnetic to denote the
non-linear response of the medium to the external field, what we have
considered actually in our previous calculations is the occurrence of
relativistic Pauli's paramagnetism. We have not fully considered the
spin-spin coupling, which would lead more definitely to Heisenberg
ferromagnetism, and is physically reasonable to expect its appearance
in nuclear matter for densities high enough. This problem has been
studied in \cite{Maruyama} via the interaction through axial vector and
tensor exchange channels. These authors show that if the interactions
are strong enough and differ in sign, the system loses the spherical
symmetry due to a mechanism independent of the one discussed in the
present paper. In the case of a quark liquid \cite{Tatsumi} the problem
has been studied under one gluon exchange interaction, and conditions
for ferromagnetism may arise.

Although the problem requires further research, one expects that if
Heisenberg ferromagnetism is to appear, it would increase the
magnetization to ${\cal M }_s=\kappa {\cal M}$, where $\kappa $ is the
internal field parameter. If $\kappa >>1$ our previous estimates for
the vanishing of the transverse pressure might be  largely exceeded,
with the arising of a new spontaneous magnetization ${\cal M}_s \sim \kappa
{\cal M}=B/4\pi$. This would mean that the magnetic field $B$ could be
kept self-consistently and our previous calculations would be a lower
bound of the fully ferromagnetic case. The vanishing of $P_{\perp}$ is
expected to occur surely at values of $B$ smaller than those depicted
in Figures 2 and 3. As a rough estimation, if we assume the exchange
interaction $J$ among neutrons of order of few hundreds of MeV, and the
number of nearest neighbors as $z \sim 10$, by dividing their product
by the dipole interaction energy, say, for the core of the star where
$N=N_0$, we get $\kappa \sim 10^4$. This means $B \sim 4\pi {\cal M}_s
\sim 10^{20}$G. For such extremely large fields the magnetic coupling
of quarks with $B$ would become of the order of their binding energy
through the color field producing a deconfinment phase transition
leading to a quark (q)-star, a pressure-induced transition to uds-quark
matter via ud-quark condensates, as discussed in
Refs.\cite{madsen99,glendenning}. But fields of that order lead surely
to the collapse of the star and even to the instability of vacuum.

\begin{figure}
\begin{center}
\end{center}
\caption{The curve $P_{\perp} = 0$ in terms of the neutron star
mean density N and the surface magnetic field B.}
\end{figure}

\section{Magnetar Formation and Stability}

Next we briefly review the basic ideas supporting the theory of
magnetars and then show why these hypothetical objects cannot
survive after reaching the claimed super strong magnetic fields.
We then present prospectives for a hybrid or strange star to
appear as a remnant of the quantum magnetic collapse of a NS.
According to Duncan and Thompson\cite{DT92}, NSs with very high
dipole surface magnetic field strength, $ B\sim [10^{14}-10^{15}]$ G, may
form when (classical) conditions for a helical dynamo action are
efficiently met during the seconds following the core-collapse and
bounce in
a supernova (SN) explosion\cite{DT92}. A newly-born NS may undergo
vigorous convection during the first 30 s following its formation
\cite{janka96}. If the NS spins (differentially) sufficiently fast
($P\sim 1$ ms) the conditions are created for the $\alpha -\Omega
$ dynamo action to be built, which may survive depletion due to
turbulent diffusion. Collapse theory, on the other hand, shows
that some pre-supernova stellar cores could adquire enough spin so
as to rotate near their Keplerian equatorial velocity, the
break-up spin frequency: $ \Omega_K \geq ([\frac 23]^3 G_N M/R^3)^{1/2}$,
which implies a period $P_K \sim 0.6$~s, after core bounce. Under these conditions, fields as large as \cite{kluzniak98,DT92}

\begin{equation}
B\sim 10^{17}\left(\frac P{1{\rm ms}}\right)\;\rm G,
\end{equation}

may be generated as long as the differential rotation is dragged
out by the growing magnetic stresses. For
this process to efficiently operate the ratio between the spin
rate ($P$) and the convection overturn time scale ($\tau _{con}$),
the Rossby number ($R_0$), should be $R_0 \leq 1$. Duncan and
Thompson warned that $ R_0 \gg 1$ should
induce less effective mean-dynamos\cite{DT92}). In this case, an
ordinary dipole $B_{D}\sim [10^{12}-10^{13}]$G may be built by
incoherent superposition of small dipoles of characteristic size
$\lambda \sim [\frac 13-1]$km, and a saturation strength
$B_{sat} = (4\pi \rho )^{1/2}\lambda /\tau _{con}\simeq 10^{16}$~G
may be reached at the surface during this early evolution of the
nascent neutron
star. At such fields, the huge rotational energy of a NS
spinning at: $ \omega_{\rm NS} \geq 1$kHz,  is leaked out via {\it
magnetic braking} and an enormous energy is injected into the SN
remnant. This process may explain the power emitted by a
{\it plerion.}

As shown above, at the end of the SN core collapse we are left with a
rapidly rotating NS endowed with with an extremely strong magnetic
field (ESMF) strength and a large matter density $\rho \sim
[10^{14}-10^{12}]$ g cm$^{-3}$. As illustrated in Figure 1, those are
the conditions for the quantum instability to start to dominate the
dynamics of the young pulsar. At this stage, the magnetic pressure
inwards may overpass the star energy density at its equator and the
collapse becomes unavoidable. As the collapse proceeds, higher and
higher densities are reached till the point the supranuclear density
may reverse the direction of implosion. A hybrid or strange star (SS)
may have  formed. We explore next this plausible outcome, among the
other possibilities quoted above. From that moment, the sound wave
generated at the
core bounce builds itself into a shock wave traveling through the star
at $ V_{SW}\sim c/\sqrt{3}$ km s$^{-1}$. Although the ESMF strength
could be quite large as long as the collapse advances, the huge kinetic
energy, $E\sim 10^{51-52}$ erg, the mean energy obtained in
calculations of energy release in neutron star phase transitions to
strange ({\it twin}) stars \cite{stoecker02,pacheco02,herman98} and
some {\it prompt
shock} supernovae ($E_{p-s} \sim 10^{51}$~erg) \cite{janka96}, carried
away by the shock wave drives a kind of supernova explosion inasmuch as
in the quark nova model and similar scenarios \cite{herman98}. Such a
huge {\it ram} pressure may counterbalance the magnetic pressure, and
even surpass it, i. e.,

\begin{equation}
\rho _{eject}V_{SW}^2\geq \frac{B^2}{8\pi \mu _0}\left( \frac
R{r_A }\right)^6,
\end{equation}

at a location from the star center equivalent to the Alfv\'{e}n
radius of the magnetar

 \begin{equation}
 r_A=\left(\frac{2\pi^2}{G\mu _0^2} \right)
^{1/7}\left[\frac{B^4R^{12}}{M\dot{M}^2}\right]^{1/7} \sim 80~ \rm km.
\end{equation}

This radius is quite large, about 7 times the NS radius (see Table 1).
Here $\dot{M}$ defines the accretion rate of the free-falling
overlaying material making up the NS crust, which is left out when the
transition occurs (see definitions and further details in
Ref.\cite{herman98}, and references therein).  Therefore, it is quite
legitimated to expect that most of the magnetic energy stored inside
the magnetosphere to be drained out of the Alfv\'en radius.  Notice in
addition that the strange star radius scales as:  $R_{\rm SS} = R_{\rm
NS} (\frac{\rho_{\rm NS}}{\rho_{\rm SS}})^{1/3}$, where the densities
ratio reads: $\frac{\rho_{\rm NS}}{\rho_{\rm SS}} \simeq 0.1-0.2$.
Other relations between both the stars can be obtained by using
conservation laws or appropriate scalings.

Then the ESMF lines are pushed out and finally broken, in a process
inverse to the standard accretion one, from $r_A$ onwards into the SN
remnant surroundings, as a violent explosion that dissipates a large
part of the magnetic flux $(\Phi \sim B^2r_A^2)$ and energy trapped
inside the magnetar magnetosphere\cite{horvath99}. Energy from the
magnetic field can be dissipated via vacuum polarization and
electron-positron pairs creation, as well as acceleration of charged
matter flowing away (synchroton and curvature losses
\cite{zhang00,zhang00A} with the explosion and material trapped in the
star magnetosphere and nearby the Alfv\'en radius. This is analogous to
the mechanism operating during a solar flare or a coronal
mass-ejection, where the very high $B$ in a given Sun-spot is
drastically diminished after flaring for a short period of time (see
also Ref.  \cite{kluzniak98}). In the Sun spots outbursts and coronal
mass ejections launch into space part of the Solar wind of charged and
neutral particles passing by the Earth\cite{parker01,chen01}. In the
case of an imploding NS, the phase of open magnetic field lines over
which the strange star is acting as a {\it propeller} lasts for about
$\Delta T_{prop} \sim E_{spin} / L_{prop} \sim [10^2-10^3]$~s, with
$L_{prop} \sim 2 \dot{M} c^2$ the propeller luminosity, and $E_{spin}
\sim I \omega^2_{\rm SS}$ the star rotational energy. Thence, the large
amount of matter ejected from the strange star at such large velocities
and the pairs created, in the vacuum breakdown, drains out the dipole
field of the remnant below the quantum electrodynamic limit of
$B_{ce}\sim 4.4 \times 10^{13}$G \cite{horvath99}.

To give an insight into this piece of the physics of
the problem, notice that once the propeller phase is over and no more
luminosity is coming from that mechanism, the magnetic field lines can
recombine  again if the energy released in this new stable
phase is essentially the strange star rotational dipole luminosity
(as the one from a millisecond pulsar), which then becomes the star
dominant mechanism of energy emission, that is

\begin{equation}
 {4} \frac{B^2_{\rm SS} R^6_{\rm SS} \omega_{\rm SS}^4}{9 c^3} =
I \omega_{\rm SS} \dot{\omega}_{\rm SS}\; .
\end{equation}

For the parameters shown in Table 1 and the observed luminosity from
fast rotating pulsars, this relation implies a new
equilibrium magnetic field $B \sim 10^{12-13}$~G, which is below the
quantum electrodynamics threshold.

Since all the differential rotation has been dragged up to build up the
former ESMF, then nothing else remains to make the magnetic field to
grow to its pre-collapse value.  Thence no such ultra high $B$ should
reappear.  We may be left with a sub-millisecond strange star\cite
{madsen99} or a hybrid star\cite{glendenning} with "canonical'' field
strength, but no any magnetar. We note in passing that the above
theoretical result is attained on the standardized assumption that the
structure of the magnetic field of the pulsar is dipolar. This premise
is underlying to the claim by Kouveliotou et al.\cite{chryssa99} that
a magnetar had been discovered in the soft gamma-ray repeater source
SGR 1806-20. Notwithstanding, for other NS (multipolar or uniform)
field configurations we do expect the overall behavior here discussed
to persist, since once the spin-spin coupling is taken into account the
unavoidable consequence is the appearance of a ferromagnetic (axial)
configuration or structure which is dipolar in nature, and therefore
the theory propound here still holds, because the magnetic field could
then be amplified by a factor $10^4$), putting the nascent pulsar above
the threshold for stability, and the collapse ensues. Rephrasing this,
one can think of this theory as a field-configuration independent
constraint on initial NS magnetic field strengths.

In looking for other contexts far from those involving compact remnant
stars, we noticed that recent ``Tabletop Astrophysics'' experiments
performed by C. Wieman et al.\cite{wieman00} have succeeded in refining
sophisticated techniques to switch atoms in a Bose-Einstein condensate
(BEC) \cite{wieman00} from states of implosion to states of
re-expansion (or explosion). In certain cases they observed that some
collapses appear rather similar to microscopic supernovae. The initial
implosion is followed by an explosion in which atoms are ejected as a
hollow ball or in narrow jets, like in the collapse and rebound of a
exploding star that forms characteristic expanding balls or streams of
outflowing gas.  The Wieman team named the phenomenon ``Bosenovae''
because of the similarities with typical supernovae. In fact, in both
phenomena some material is left after the implosion as a compact
object. Because of the similarity of the physics of Bose-Einstein
condensates with the one we presented above, we are confident that this
our theory can also be adapted to explain these impressive results by
Wieman et al.\cite{wieman00}; Ketterle and Anglin \cite{Ketterle}.
The main feature of these oscillating Bose-Einstein condensates, said
to be a scale-down version of either a neutron star or a white
dwarf\cite{wieman00}, is that for some critical fields what appears is
an attractive force between the atoms and the condensate implodes and
rebounds driven by some sort of internal negative pressure.  We claim
in this paper (a detailed description of BECs phenomenology is to be
given in a work in preparation\cite{nos2003}) that such a negative
pressure could be explained in the context of the theory introduced
in this paper, since a precise relationship between gas density and
magnetic
field strength in the BEC is settled out by switching the atoms between
attractive and non-attractive states. Thence the negative pressure acts
as the equivalent of an attractive force among the atoms directed
towards the magnetic field axis, leading to the BEC implosion.

\section{Conclusions}

We conclude by claiming that if a degenerate neutron gas is under the
action of a super strong magnetic field $B_{ce} \lesssim B \lesssim
B_{cn}$, for values of the density typical of NS matter its transverse
pressure vanishes, the outcome being a transverse collapse. This
phenomenon establishes a strong bound on the magnetic field strength
expected to be found in any stable neutron star pulsar, regardless of
its initial field configuration,  and suggests a
possible endpoint in the early evolution of highly magnetized neutron
stars.  They could likely be a mixed phase of nucleons and a $\pi^{\pm},
\pi^0, K^{\pm }, K^0, \bar{K}^0, \sigma, \rho^{\pm }, \omega $ meson
condensate, a hybrid or strange star, or a distorted black hole but no
any magnetar at all. We point out, nonetheless, that if by any mechanism
a strange star could be formed directly in a supernova exlposion (which is
uncertain, but not ruled out) and if vigorous dynamo action operates in
the strange quark matter bulk, "pulsars" with fields higher than $B_{ce}$
could still be formed. In other words, if any magnetars exist, they cannot
be neutron stars.

\section{Appendix}

We use the integral representation

\begin{equation}
a^{1/2}=\pi^{-1/2}\int_0^{\infty}\frac{dx}{x^2}(e^{-ax^2}-1)=
\pi^{-1/2}\int_0^{\infty}dy y^{-3/2} (e^{-ay}-1) \label{Ar} .
\end{equation}

We regularize   the divergent term dependent on $a$ in (\ref{Ar}),
by introducing a small quantity $\epsilon$ as the lower limit in
the integral, and neglecting the term independent of $a$

\begin{equation}
a^{1/2}(\epsilon)=\pi^{-1/2}\int_\epsilon^{\infty}dy y^{-3/2}
e^{-ay} \label{Ar1}.
\end{equation}

By taking $a(\epsilon,\eta)=p_3^2 + (\sqrt{p_{\perp}^2+m_n^2}+\eta
qB)^2$ and substituting in (\ref{ONT1}) and performing the
Gaussian integral on $p_3$, one obtains

\begin{equation}
\Omega_{Vn}(\epsilon)=\frac{1}{4\pi^2}\sum_{\eta=\pm 1
}\int_0^\infty p_{\perp}dp_{\perp} \int_\epsilon^{\infty}dy
y^{-5/2} e^{-(\sqrt{p_{\perp}^2+m_n^2}+\eta qB)^2 y}. \label{Ar2}
\end{equation}

\noindent
By substituting $z= \sqrt{p_{\perp}^2+m_n^2} +\eta qB$
one is left with the expression

\begin{eqnarray}
\Omega_{Vn}(\epsilon) & = & \frac{1}{4\pi^2}\int_\epsilon^{\infty}dy
y^{-3} e^{-(m_n^2+q^2B^2) y} \cosh( q B m y) \nonumber  \\
& + & \sum_{\eta=\pm 1}\frac{\eta q B}{4\pi^2}\int_\epsilon^{\infty}dy
y^{-2}\int_{m_n + \eta q B}^\infty e^{-z^2}y dz. \label{Ar3}
\end{eqnarray}

By introducing the new variable $w = z-m_n -\eta qB$, the second
integral in (\ref{Ar3}) becomes

\begin{equation}
\frac{\eta q B}{4\pi^2}\int_\epsilon^{\infty}dy
y^{-2}\int_0^\infty e^{-[(w+m_n)^2+ q^2
B^2]y}\sinh(2qB(w+m_n)y)\label{Ar4}.
\end{equation}

After subtracting to $\cosh(qBmy)$ and $\sinh[2qB(w+m_n)y]$ the
first two terms in their series expansion, one can take $\epsilon
\to 0$ and obtain the finite expression (\ref{nvac}). This process
is equivalent to the subtraction of divergent terms, one of which
is proportional to $B^2$, and absorbs the classical field energy
term $B^2/8\pi$.

It is not
difficult to check that for fields $B \ll 10^{20}$ G, the first
term in (\ref{nvac}) is the dominant one. Its first contribution
after the series expansion of $\cosh(qBmy)$ is
$q^3B^3m_n^3/2\pi^2(m_n^2+q^2B^2)$. For fields of order $10^{17}$G
such a term is of order $10^{30}$ ergs/cm$^3$, much smaller than
$\Omega_{sn}$. But for fields near $10^{20}$G its contribution is
comparable to that of $\Omega_{sn}$.

\section{Acknowledgement}
 All authors thank J. Arponen, M. Chaichian, J. Ellis, A. Green,
K. Kajantie, C. Montonen, A.E. Shabad, and A. Zepeda for useful
comments and suggestions. H.P.R thanks M. Virasoro, IAEA and
UNESCO for hospitality in ICTP. The financial support of the
Academy of Finland under the project No. 163394 is greatly
acknowledged. HJMC thanks Funda\c c$\tilde{a}$o de Amparo \'a
Pesquisa do Estado de Rio de Janeiro (Brasil) for a Grant-in-Aid.
The authors thank the anonimous referee for his many valuable criticisms
and suggestions, which led to this highly improved version of our former
manuscript.

\begin{table*}[bth]
\caption{\label{tbl-2}{Parameters used in modeling the phase transition of a neutron to a strange star, as discussed in the text. } }
\begin{tabular}{ccccccc}
\multicolumn{1}{c}{ } &
\multicolumn{1}{c}{ } &
\multicolumn{1}{c}{{} } &
\multicolumn{1}{c}{ } &
\multicolumn{1}{c}{ } &
\multicolumn{1}{c} { } &
\multicolumn{1}{c} { } \\
\vspace{3pt}
{  } & { massa M  } & { radius R } & { period P } & { magn.
field B } & { accretion rate } & {       } \\
\vspace{3pt}
{  } & { [ M$_\odot$ ]  } & { [ km ] } & {  [ ms ] } & { [ G ] } & { [ M$_\odot$~s$^{-1}$ ]  } & {       } \\
{ Neutron star  } & { 1.5 } & { 12.5 } & { 2.0 } & { $2\times 10^{15}$ }
& { $10^{-15}$ } & {       } \\
\\
{ Strange star  } & { 1.5   } & { 9.5-10.0 } & { 0.5 } & { $2\times 10^{13}$
} & { $10^{-5}$ } & {      }
\\
\end{tabular}
\end{table*}

\nopagebreak


\begin{references}

\bibitem{Chaichian}  M. Chaichian, S. Masood, C. Montonen, A. P\'{e}rez
Mart\'{i}nez, H. P\'{e}rez Rojas, Phys. Rev. Lett. {84}, 5261
(2000).

\bibitem{Hugo}A. P\'{e}rez Mart\'{i}nez, H. P\'{e}rez Rojas, H.
Mosquera Cuesta, hep-ph/0011399

\bibitem{DT92}  R. C. Duncan, C. Thompson, Ap. J. 392, L9 (1992).

\bibitem{Khalilov} V. R. Khalilov, Phys. Rev D, 65, 056001-1
(2002)

\bibitem{Chakrabarty1} S. Ghosh, S. Mandal, S. Chakrabarty,
astro-ph/0207492 (2002).

\bibitem{Herrera}  L. Herrera and N.O. Santos, Local Anisotropy in
Self-Gravitating Systems, Phys. Rep. 286, 53 (1997)

\bibitem{Mak} M.K. Mak and T. Harko, gr-qc/0110103.

\bibitem{Sommerfeld} A. Sommerfeld, Electrodynamics, Academic
Press, New York (1952)

\bibitem{Jackson}  J.D. Jackson, Classical Electrodynamics, J. Wiley
and Sons, New York, (1966).

\bibitem{Landau3}  L. D. Landau, E. M. Lifshitz, The Classical Theory
of Fields, Pergamon, New York, (1970)

\bibitem{Shabad} A.E. Shabad, H. Perez Rojas, Ann. of Phys., 121 (1979)
432.


\bibitem{haensel-bonazzola96}P. Haensel, S. Bonazzola, Astron.
Astrophys. 314, 1017 (1996).

\bibitem{horvath98}J. E. Horvath, J. A. de Freitas Pacheco , Int.
Jour. Mod. Phys. D 8, 19 (1998).

\bibitem{Ternov}  V.G. Bagrov, D.M. Gitman, Exact Solutions of
Relativistic Wave Equations, Kluwer Acad. Publ. (1990).

\bibitem{Euler} W. Heisenberg and H. Euler, Z. Phys. 98, 714
(1936).

\bibitem{Landau0}  L. D. Landau, E. M. Lifshitz, Statistical Physics,
Pergamon, Oxford York, (1958)

\bibitem{Huang} K. Huang, Statistical Mechanics, J. Wiley and Sons,
New York, (1963)

\bibitem{Weinberg} S. Weinberg, Gravitation and Cosmology, J.
Wiley and Sons, New York, (1972)

\bibitem{Landau}  L. D. Landau, E. M. Lifshitz, Electrodynamics of
Continuous Media, Pergamon, New York, (1963)

\bibitem{nos04} H. J. Mosquera Cuesta, H. P\'erez Rojas, A.
P\'erez Mart\'{\i}nez, \textit{Anisotropic pressures in very dense
matter}, in preparation


\bibitem{Konno} K. Konno, T. Obata, Y. Kojima, Astron. Astrphys. 356, 234
(2000)

\bibitem{Cardall}  C. Y. Cardall, M. Prakash, J. M. Lattimer,
Astrophys. J. 554, 322 (2001).

\bibitem{Chakrabarty}  S. Chakrabarty, D. Bandyopadhyay, S. Pal, Phys.
Rev. Lett. 78, 2898 (1997).

\bibitem{Broderick}  A. Broderick, M. Prakash, J. M. Lattimer, Ap. J.
537, 351 (2000).

\bibitem{Chiba}  T. Chiba, Prog. Theor. Phys. 95, 321 (1996).


356, 234 (2000).

\bibitem{Ketterle} W. Ketterle and J. Anglin, Bose-Einstein Condensation
of Atomic Gases, Nature 216, 211 (2002).

\bibitem{Kondratyev02} V.N. Kondratyev, Phys. Rev. Lett. 88, 221101 (2002)

\bibitem{horvath99}J. E. Horvath, Int. Jour. Mod. Phys. D 8, 669
(1999).

\bibitem{Fradkin} E. S. Fradkin, Quantum Field Theory and
Hydrodynamics, Proc. of the P.N. Lebedev Inst. No. 29, Consultants
Bureau (1967).

\bibitem{Weisskopf} V. Weisskopf, Kong. Dans. Videns. Selskab,
Math-fys. Meddeltser,14, (1936).

\bibitem{Hugo1} H. Perez Rojas, Acta Phys. Pol. B17, 861 (1986).

\bibitem{Mathews}  I-S. Suh, G. J. Mathews, Ap. J. 530, 949 (2000).

\bibitem{Chakrabarty2} S. Mandal, S. Chakrabarty,
astro-ph/0209462 (2002).

\bibitem{Maruyama} T. Maruyama and T. Tatsumi, Nucl. Phys. A 693, 710
(2001)
\bibitem{Tatsumi} T. Tatsumi, Phys. Lett. B 489, 310 (2000)

\bibitem{stoecker02}I. N. Mishustin, et al., Phys. Lett. B 552, 1
(2003).

\bibitem{pacheco02}G. Marranghello, C. Z. Vasconzellos and J. A. de
Freitas Pacheco, Phys. Rev. D 66, 064027 (2002).

\bibitem{herman98}  H. J. Mosquera Cuesta, Ph. D. thesis (in
portuguese), INPE, Brazil, unpublished (1998). See also the paper in
preparation: "Gravitational waves, gamma-ray bursts and ultra high
energy cosmic ray acceleration during neutron star phase
transitions: An unified picture", by H. J. Mosquera Cuesta, J.
A. de Freitas Pacheco and M. Dillig. The "Quark Nova" model can
be found in R. Ouyed, J. Dey, M. Dey, Astron. Astrophys. 390,
L39 (2002).


\bibitem{janka96}  H.-Th. Janka, E. M\"{u}ller, A. \& A. 306, 167
(1996).

\bibitem{zhang00} B. Zhang and A. K. Harding, Astrophys. J. 535, L51 (2000).


\bibitem{zhang00A} B. Zhang, A. K. Harding and A. G. Muslimov, Astrophys.
J. 531, L135 (2000).

\bibitem{kluzniak98}  W. Klu\'{z}niak, M. Ruderman, Ap. J. Lett. 505,
L113 (1998). See also M. Ruderman, L. Tao, W. Klu\'{z}niak, Ap. J.
542, 243 (2001).

\bibitem{madsen99}  J. Madsen, Phys. Rev. Lett. 81, 3311 (1998).

\bibitem{glendenning}  N. K. Glendenning, {\it Compact stars: Nuclear
physics, particle physics and general relativity}, New York, USA:
Springer (1997).


\bibitem{chryssa99}C. Kouveliotou, et al., Nature 393, 235 (1998),
and also C. Kouveliotou, et al., Astrophys. J. 510, L115 (1999).









\bibitem{parker01}E. N. Parker, Chin. Journ. Astron. Astrophys. 1, 99 (2001).


\bibitem{chen01}J. Chen, Spac. Sci. Rev. Vol.95, issue 1/2, 165 (2001).


\bibitem{wieman00}S. L. Cornish, et al., Phys. Rev. Lett. 85, 1795 (2000).

\bibitem{nos2003} H. J. Mosquera Cuesta, H. P\'erez Rojas, A.
P\'erez Mart\'{\i}nez, \textit{Exploding Bose-Einstein condensate
and critical magnetic fields}, in preparation



\end{references}
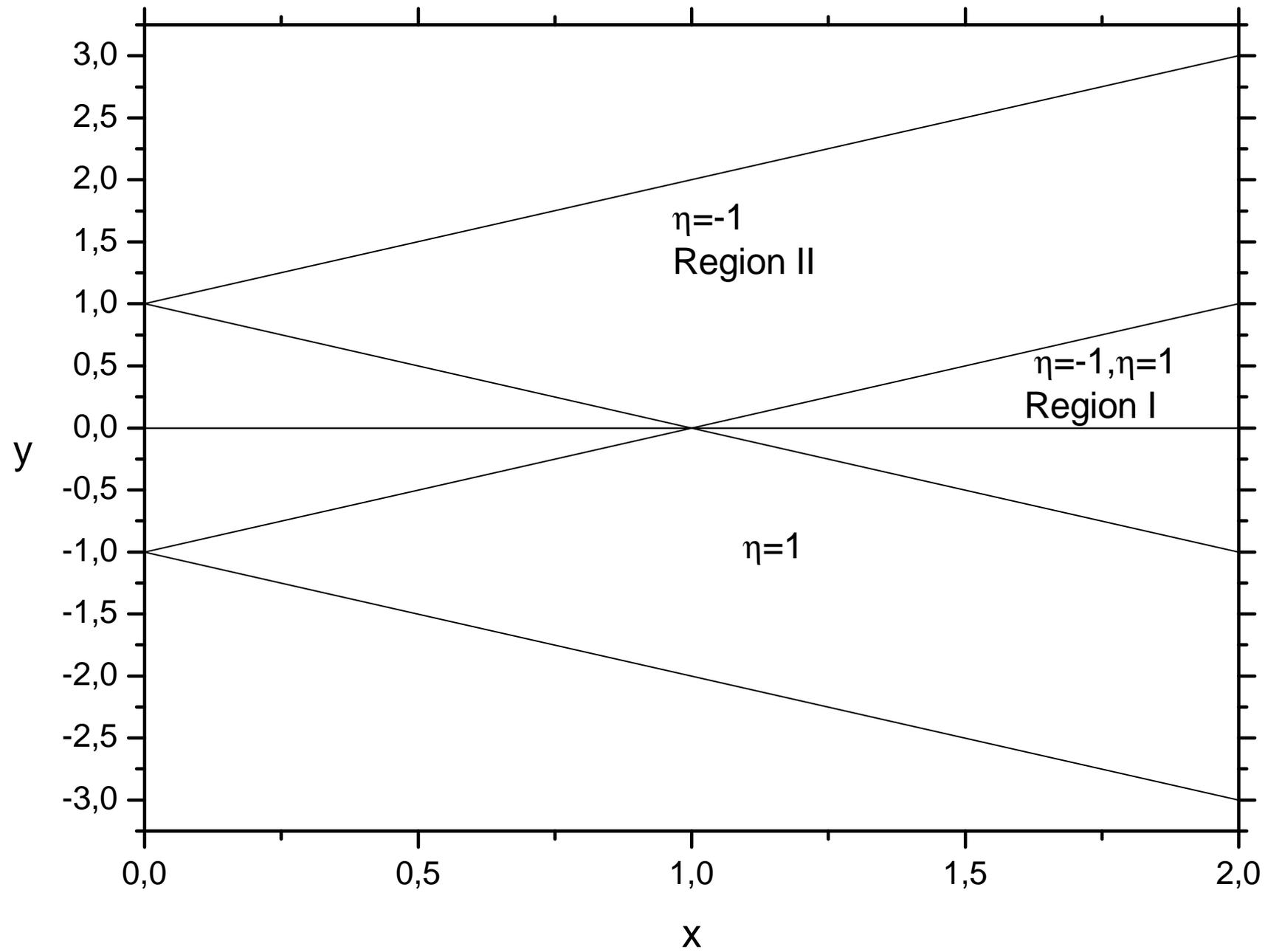
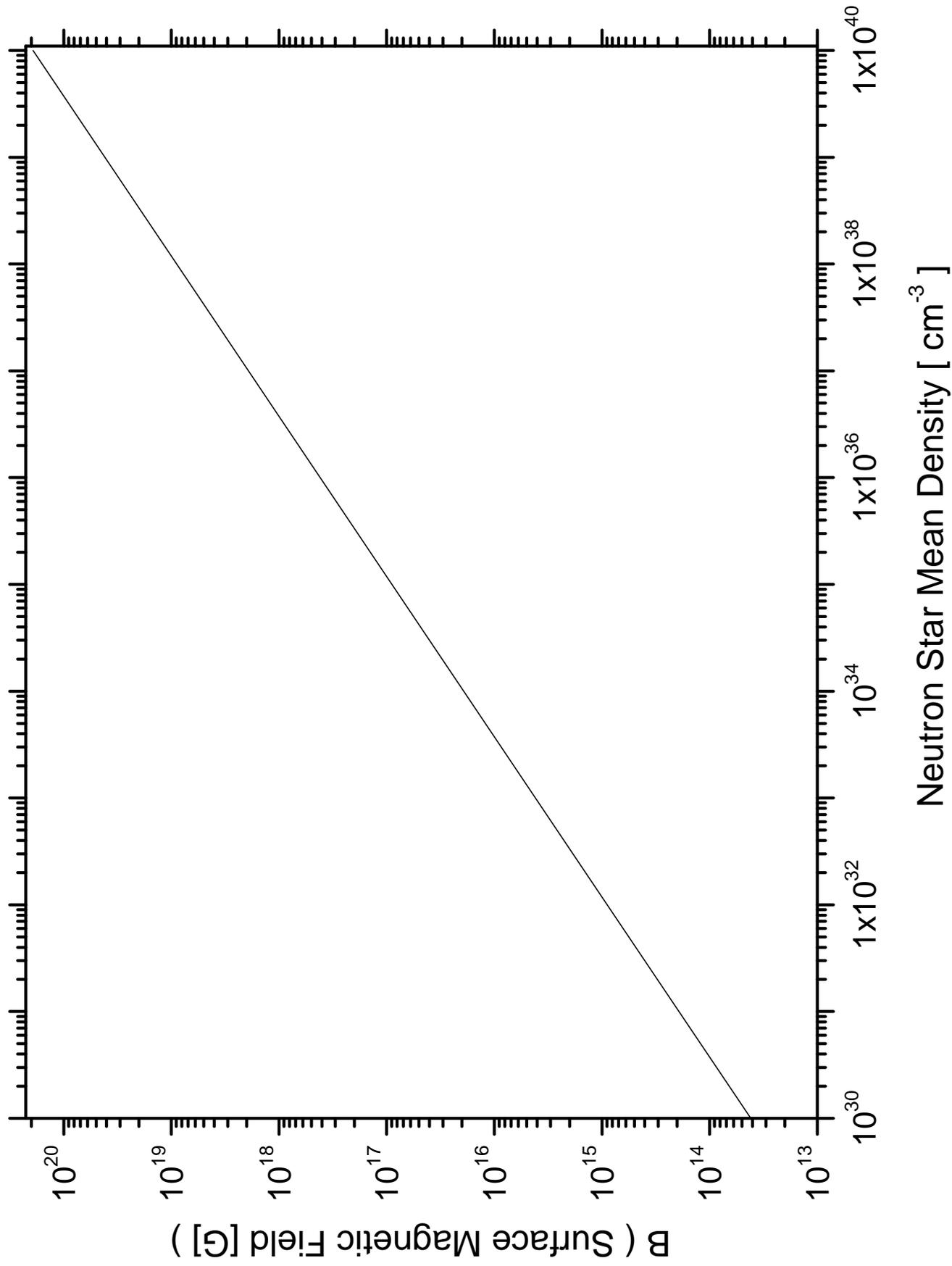
\end{document}